\documentclass[12pt,osajnl2,preprint,showpacs,superscriptaddress]{revtex4}  
\usepackage[draft]{hyperref} 

\begin{document}

\title{The Effect of Immersion Oil in the Optical Tweezers}

\author{Ali Mahmoudi, and S. Nader S. Reihani}
\address{Department of Physics, Institute for Advanced Studies in Basic Sciences
(IASBS), Gava Zang, P.O.Box:45195-1159, Zanjan 45137-66731, Iran.}
\email{}

\begin{abstract}
In this paper, we present a theoretical study on the effect of
refractive index of immersion oil on the position of optimal depth
and optical trap quality in optical tweezers. Using simple
numerical calculation presented here, one can study the optical
trapping in a realistic setup. The electric field and intensity
distribution in sample medium is derived. Our calculations is in
very good agreement with the previous reported experimental results.
\end{abstract}


\maketitle

\section{Introduction}
Optical Tweezers are used as micromanipulation tools in many
scientific areas, from biology\cite{ashkin2, Bustamante1, Block1,
Hansen1}  to nanotechnology\cite{Agarwal1, shida1, Bosanac1,
seol1}. A typical Optical tweezer (OT) consists of a Gaussian
laser beam tightly focused through a high Numerical Aperture (NA)
objective lens producing a 3-D Gaussian intensity profile at the
focus. An object with the refractive index greater than that of
the surrounding medium experiences a Hookean force towards the
focus\cite{ashkin1}. The strength of the trap can be regarded as
the spring constant. OTs are widely used as non-contact
micromanipulator using a micron (and nano)-sized object as a
handle. For biological applications a near-infra-red laser beam is
hired to minimize the damage to the sample \cite{Block1}.
Nanometer spacial resolution along with sub-Megahertz temporal
resolution have turned OT to a widely desired tool in many
scientific communities. OT are normally implemented into an
optical microscope in order to visualize the specimen under
manipulation. It is experimentally shown that refractive index
discontinuity in the optical pathway of OT will introduce
considerable amount of spherical aberrations. To compensate for
such an aberration, different methods are
proposed\cite{Reihani-oil, mingu, Reihani-tube, deformable} among
which the changing the refractive index of the immersion
medium\cite{Reihani-oil} seems to be more feasible. In this letter
we present a theoretical calculation which gives rise to the
refractive index of the immersion medium at which the optimal trap
occurs. Trapping in water and air are considered as examples. The
theoretical results are in very good agreement with the previously
available experimental results.

\section{The effect of immersion oil on the intensity distribution in focal region }
If the image space in an aplanatic system is homogeneous, then
electric field at point P  located around the focus(origin located
at the focus) can be written as\cite{wolf}:
 \begin{eqnarray}
\vec{E}(P)=-\frac{ik_1}{2\pi}\int\int_{\Omega}\frac{\vec{a}(s_x,s_y)}{s_z}\exp \left ( ik[\Phi(s_x,s_y)+\hat{s}.\vec{r}_P]\right )ds_{1x}ds_{1y}
\end{eqnarray}
Where k is wavenumber, $\hat{s}=s_x\hat{i}+s_y\hat{j}+s_z\hat{k}$
is a unit vector along a typical ray,$\Omega$ denotes the lens
aperture, $\vec{r}_P$ is position vector of point P,$\Phi$ is
aberration function of the lens and finally $\vec{a}$ is the
electric strength vector. A similar equation can be written for
magnetic field. It can be shown that if there has been a
refractive index mismatch in image space, namely we have a planar
interface between two media with refractive indices $n_1$ and
$n_2$, then assuming the lens is aberration free(or has a constant
aberration) the electric field on this boundary($z=-z_I$) and on
the side of second medium can be written as\cite{torok}:
\begin{eqnarray}
\vec{E}_2(x,y,-z_I)=-\frac{ik_1}{2\pi}\int\int_{\Omega}{\bf T}^{1\rightarrow 2}{\bf W}_{e}(\hat{s}_1)\exp \left ( ik_1(s_{1x}x+s_{1y}y-s_{1z}z_I)\right )ds_{1x}ds_{1y}
\end{eqnarray}
Where $k_1$ is wavenumber in the first medium, $\Omega$ is
denoting the surface of objective lens,${\bf
W}_{e}=\frac{\vec{a}(s_{1x},s_{1y}}{s_{1z}}$ and ${\bf
T}^{1\rightarrow 2}$ is an operator that describes the changes in
electric field on crossing the boundary and it is a function of
$n_1,n_2$ and incidence and refraction angles at this interface\cite{torok}.
Equation(2) can be extended to the general case when there are m
different media with m-1 interfaces. Assuming that the refractive
indices of these media are $n_1,...,n_m$, and a linear polarized
Gaussian beam $\vec{E}=E_0e^{-\rho^2/w_0^2}\hat{i}$ ($w_0$ is the beam
waist and $\rho=\sqrt{x^2+y^2}$)incident on the front aperture of
the objective lens, it can be shown that the electric field inside the
m-th medium after transforming to spherical coordinates can be written as:
\begin{eqnarray}
\vec{E}_m(x,y,z) & = & -\frac{ik_1}{2\pi}E_0\sqrt{\frac{n_1}{n_2}}\int_{0}^{\alpha}\int_{0}^{2\pi}\frac{E_{sample}}{s_{1z}}\\ & & \nonumber\exp \left ( ik_0\left[n_1(t_2+t_3+...+t_m)\cos\phi_1-  n_2t_2\cos\phi_2-...-n_mt_m\cos\phi_m \right]\right ) \\ & & \nonumber \exp\left[in_mk_0z\cos\phi_m\right]\exp\left[in_1k_0\sin\phi_1(x\cos\theta+y\sin\theta) \right]\sin\phi_1\cos\phi_1^{1/2}d\theta d\phi_1
\end{eqnarray}
Where $\alpha$ is the convergence angle of the objective given by
the Numerical Aperture of  the objective ($NA=n\sin\alpha$, with n
being the refractive index of the immersion medium),$t_k
(k=2,3,...,m$  is the thickness of k-th medium, and the angles
$\phi_i,(i=2...m)$ are refraction angles in different media.
$E_{sample}$ is the electric field strength vector inside the
sample medium given by $E_{sample,x}=\prod_{l=1}^{l=m-1}\tau_{pl}\cos^2\theta\cos\phi_{m}+\prod_{h=1}^{h=m-1}\tau_{sh}\sin^2\theta$
,$E_{sample,y}=\prod_{l=1}^{l=m-1}\tau_{pl}\cos\theta\sin\theta\cos\phi_{m}-\prod_{h=1}^{h=m-1}\tau_{sh} \cos\theta\sin\theta$ and $E_{sample,z}=-\prod_{l=1}^{l=m-1}\tau_{pl}\cos\theta\sin\phi_{m}$.
%
%
Normally for optical trapping applications there would be 3
discontinuity in the refractive  indices (m=4), with media being
the objective ($n_1=n_{obj}=1.518$), immersion oil ($n_2=n_{imm}$,
$t_2=Y$, Y is variable), coverglass ($n_3=n_{g}=1.518$,
$t_3=170\mu m$), and sample ($n_4=n_{s}$, $t_4=d$=probe depth).
Note that when the objective's top lens, coverglass, and the
immersion medium are index matched $(n_{obj}=n_g=1.518)$, there
would be only one interface (coverslip-water). Considering that
the restoring force of optical tweezers is proportional to the
intensity gradient\cite{akbar}, one can use equation 3 to find the
optimal parameters (such as refractive index of the immersion
medium\cite{oil}) for a desired optical trapping experiment. in
the following sections the results for the two most popular cases
will be presented.

\subsection {Trapping in water}
For the popular case of trapping inside water using an oil
immersion objective, one  can consider $n_{s}=1.33$, and
$n_{imm}=1.518$. Figure 1 shows the resulted axial intensity
profiles at different depths (a) as well as the calculated average
intensity gradients (b) for a 1$\mu m$ polystyrene bead trapped
using an objective with NA=1.3 and working distance (W.D) of
$200\mu m$ through a coverglass of $170\mu m$ thick.

\begin{figure}[h!] \centering
\includegraphics[width=0.75\textwidth,clip]{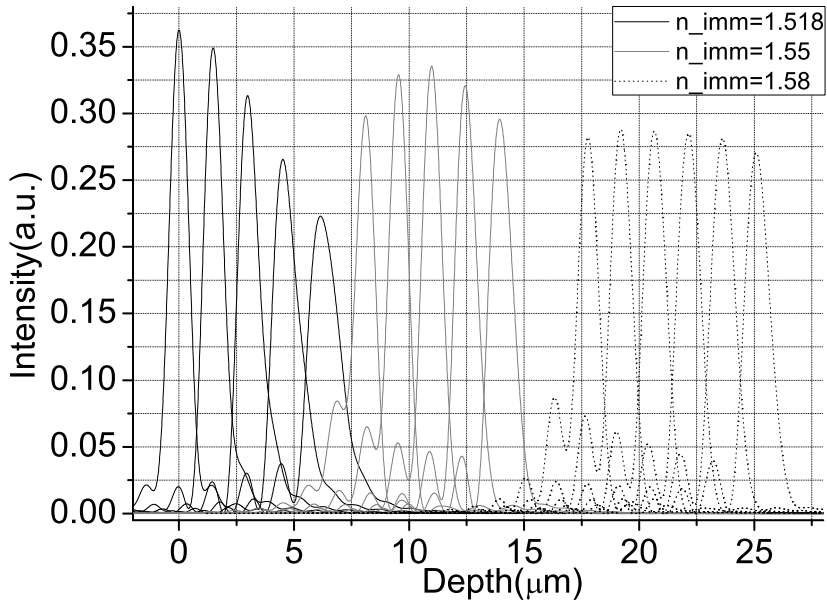}
\hfill \caption[]{ Intensity distributions in the axial direction
for different immersion oils.}
\end{figure}

\begin{figure}[h!] \centering
\includegraphics[width=0.75\textwidth,clip]{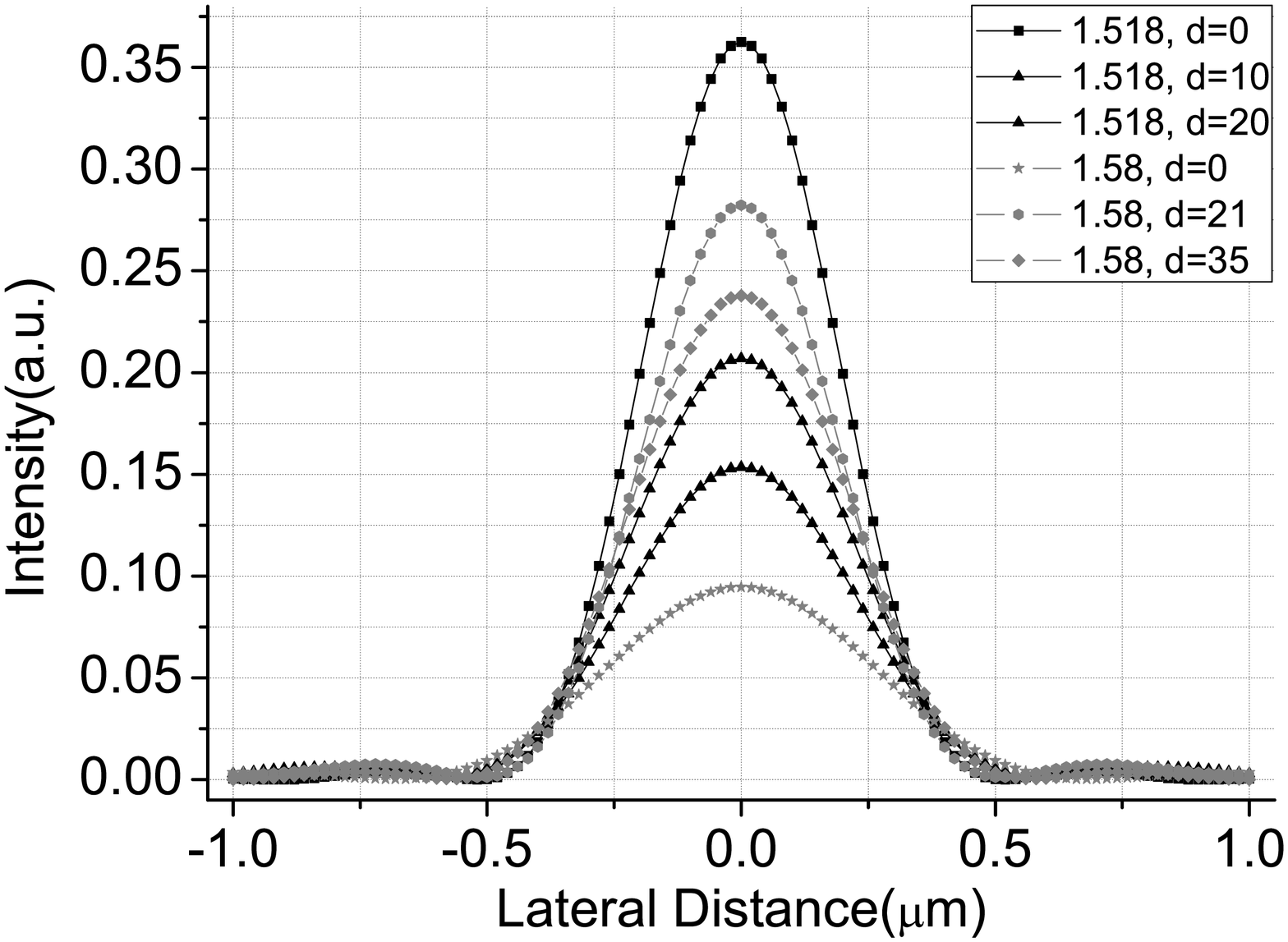}
\hfill \caption[]{ Intensity distributions in the lateral
direction for different immersion oils.}
\end{figure}

\begin{figure}[h!] \centering
\includegraphics[width=0.75\textwidth,clip]{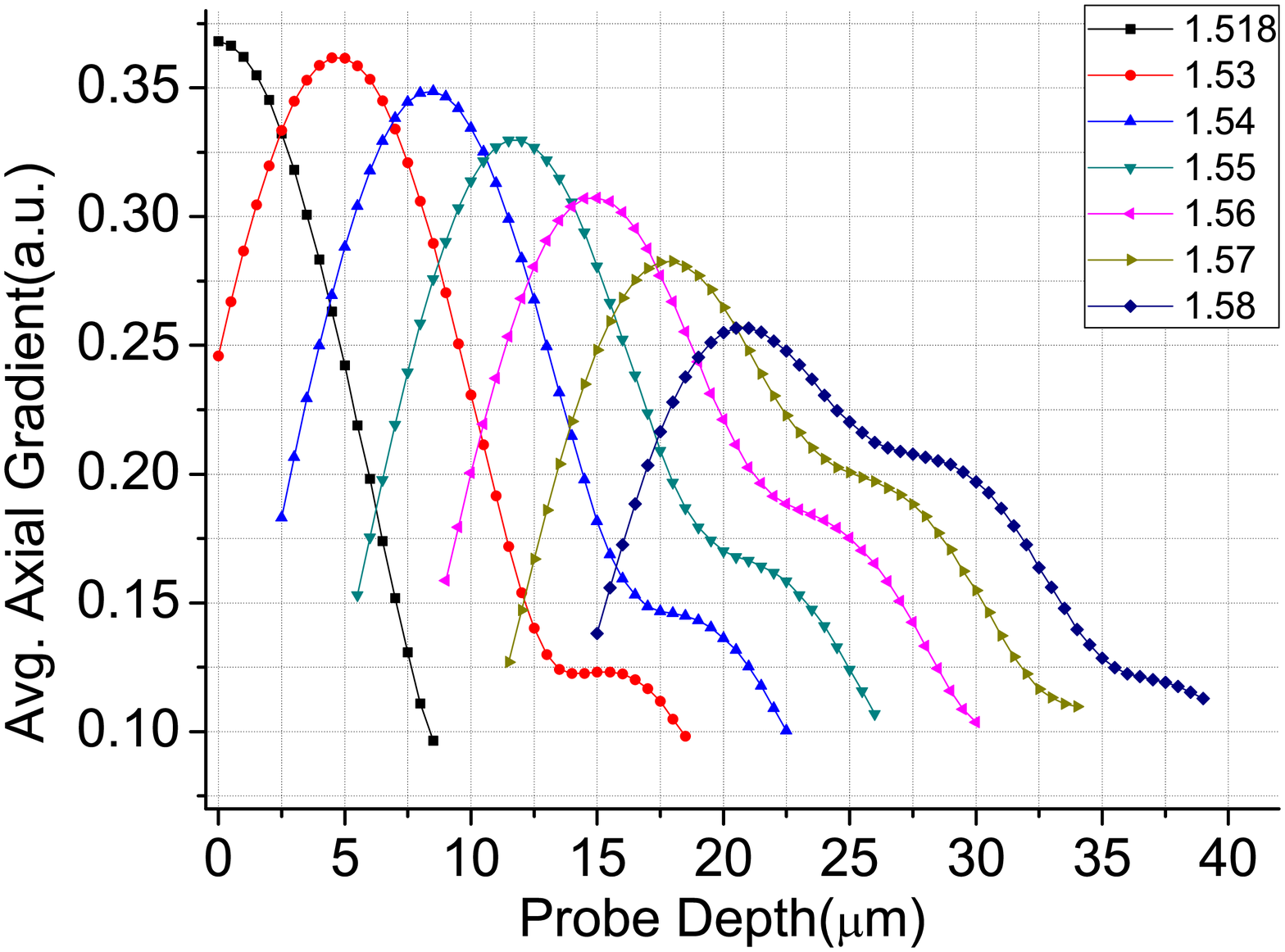}
\hfill \caption[]{ Average Intensity Gradient in the axial
direction  for different  immersion oils , trapping medium is
water.}
\end{figure}

\begin{figure}[h!] \centering
\includegraphics[width=0.75\textwidth,clip]{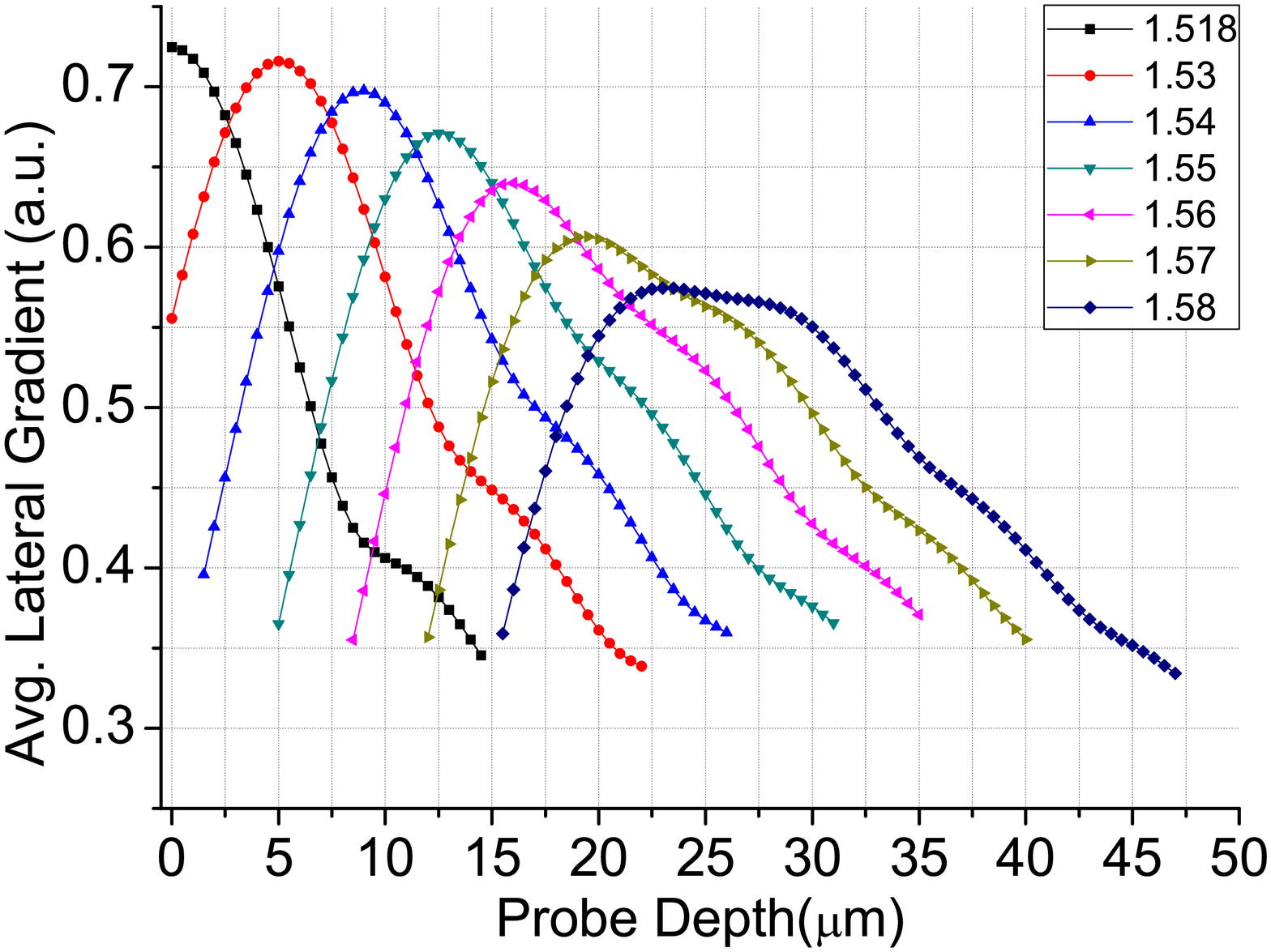}
\hfill \caption[]{ Average Intensity Gradient in the lateral
direction for different  immersion oils , trapping medium is
water.}
\end{figure}

Figure 1 illustrates that: (1) for $n=1.518$, where the system is
supposed to be  abberation-free, the optimal trap occurs just in
the vicinity of the coverglass inner surface. (2) The trapping
strength significantly decreases as the trapping depth is
increased. (3) By increasing the refractive index of the immersion
medium, the optimal trapping depth (minimum spherical aberration)
shifts towards the deeper axial positions. (4) The maximum
trapping strength decreases slightly by increasing the $n_{im}$
which implies the slight increase in the residue of spherical
aberration and reflection at boundaries. Table 1 quantitatively
summarizes the optimal conditions using different immersion oils.

\begin{figure}[h!] \centering
\includegraphics[width=0.75\textwidth,clip]{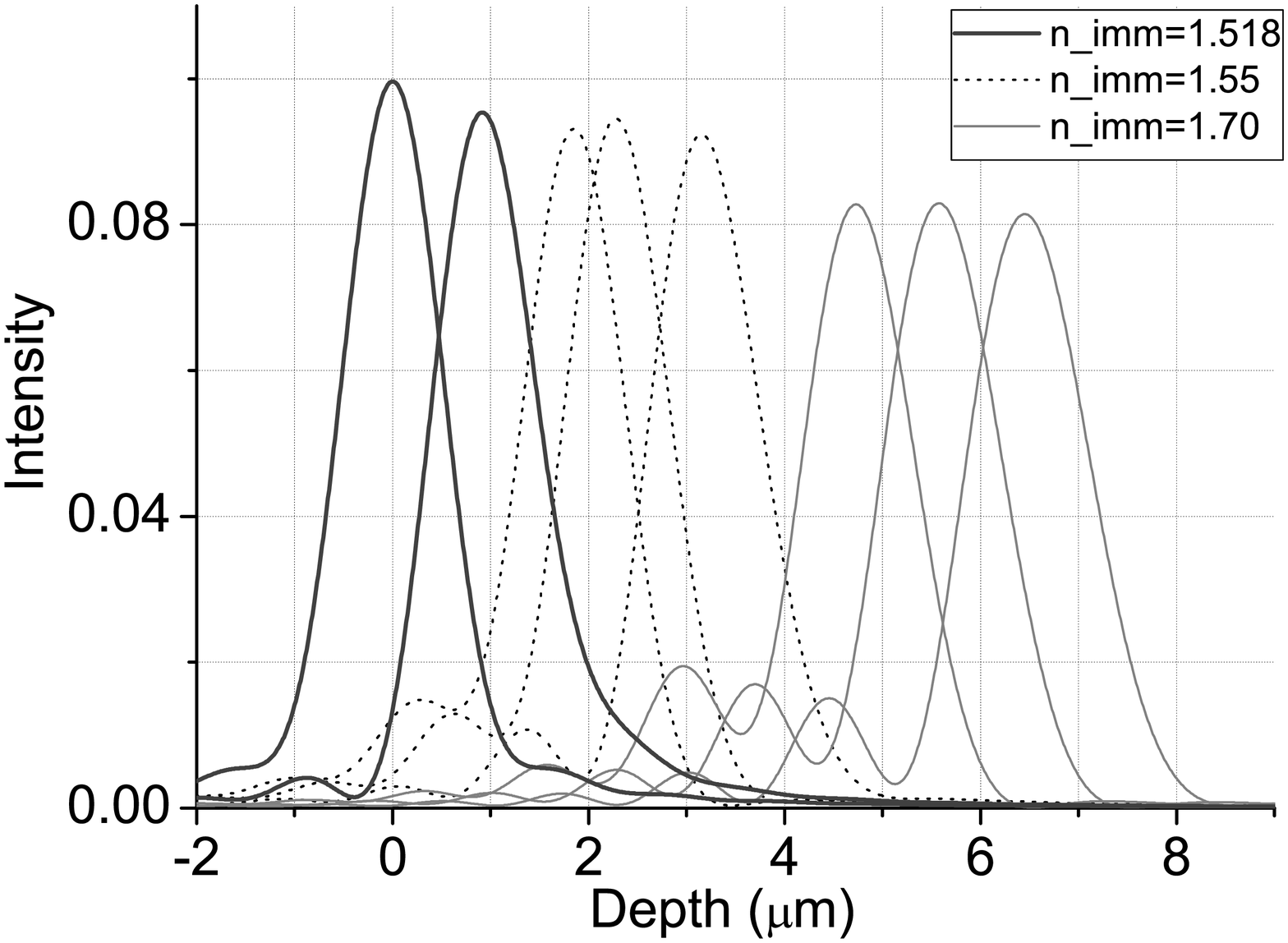}
\hfill \caption[]{ Intensity distributions in the axial direction
for three different immersion oils , trapping medium is air.}
\end{figure}

\begin{figure}[h!] \centering
\includegraphics[width=0.75\textwidth,clip]{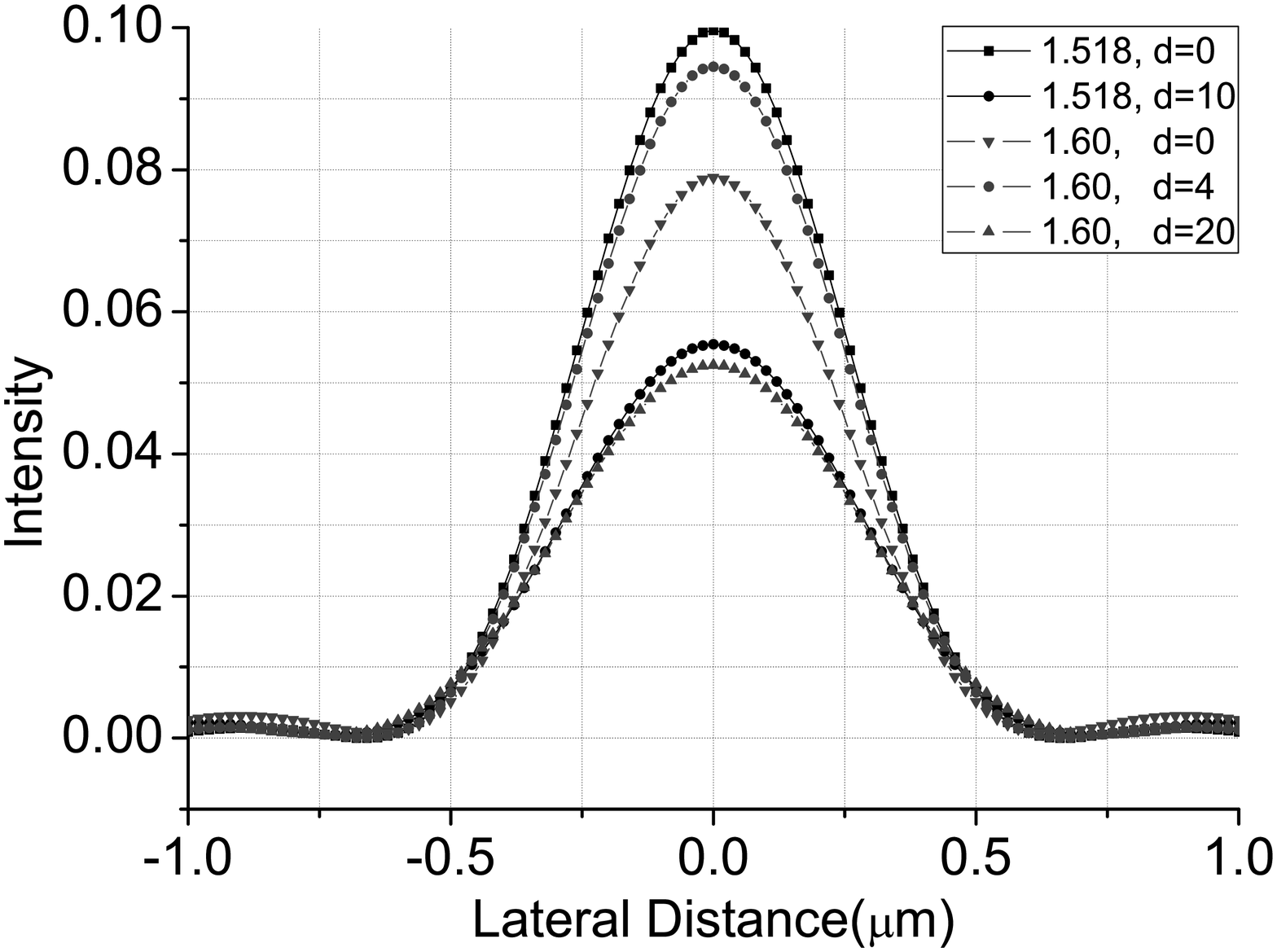}
\hfill \caption[]{ Intensity distributions in the lateral
direction for three different immersion oils , trapping medium is
air.}
\end{figure}

\begin{figure}[h!] \centering
\includegraphics[width=0.75\textwidth,clip]{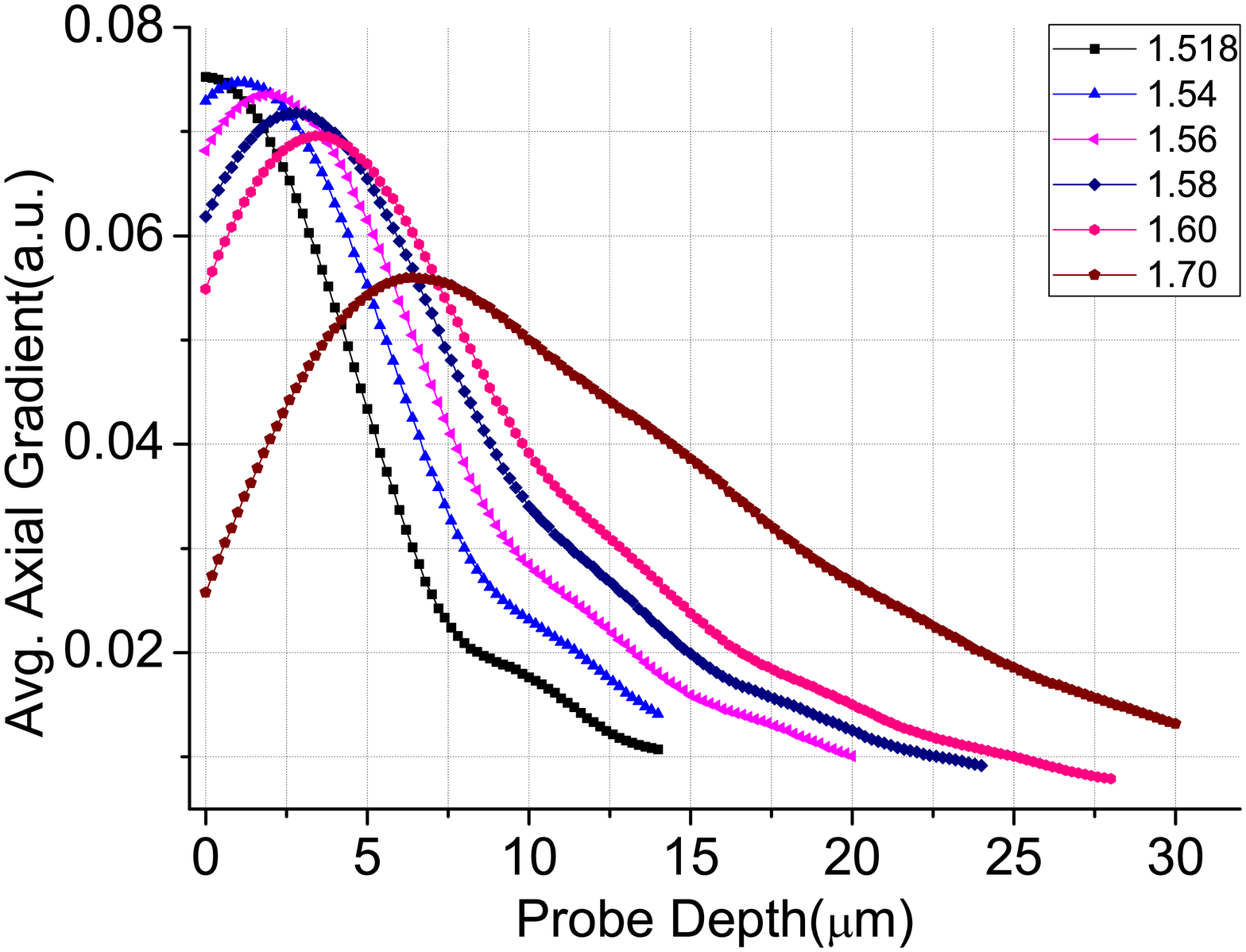}
\hfill \caption[]{ Average Intensity Gradient in the axial
direction  for different  immersion oils , trapping medium is
air.}
\end{figure}

\begin{figure}[h!] \centering
\includegraphics[width=0.75\textwidth,clip]{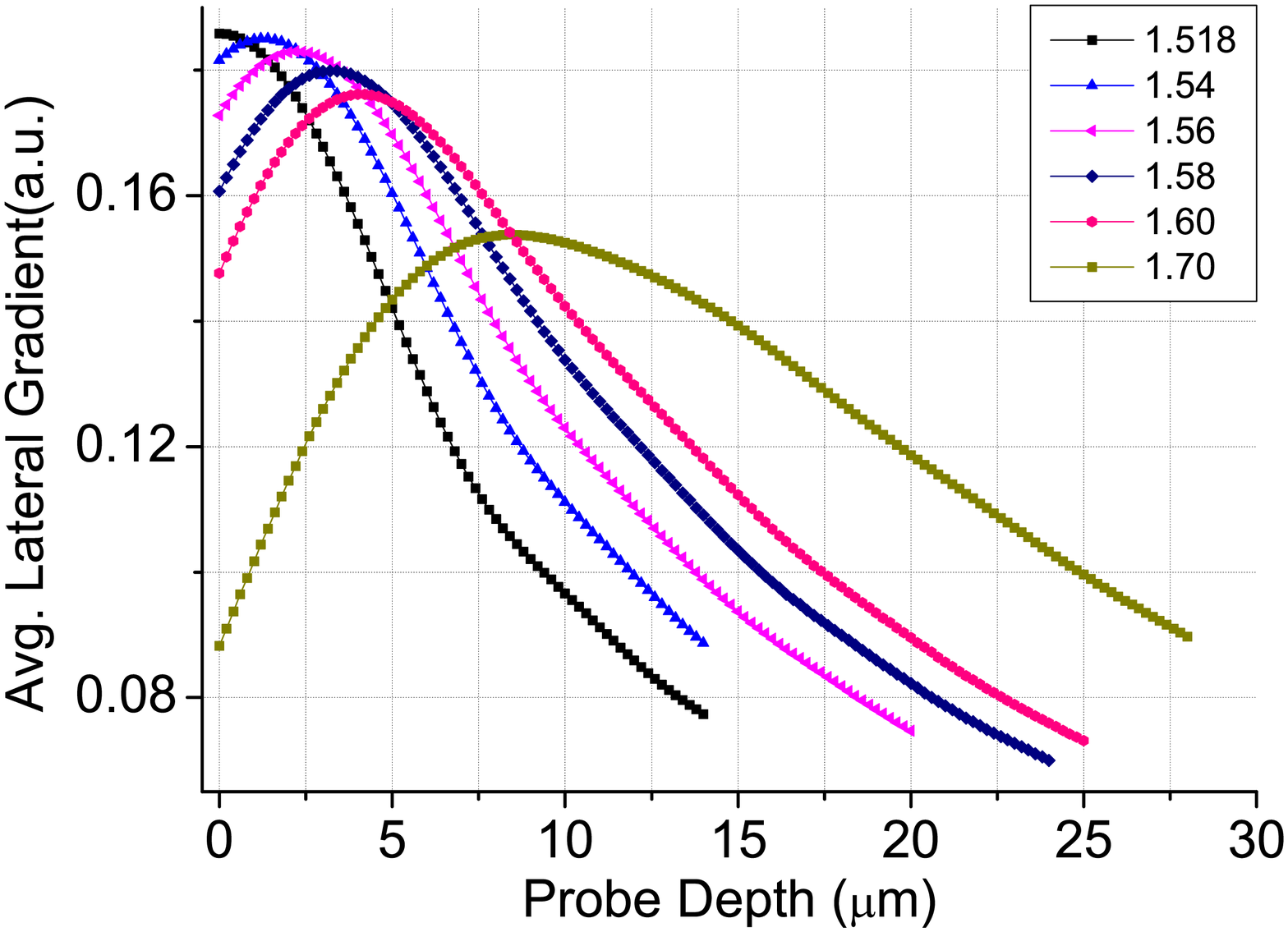}
\hfill \caption[]{ Average Intensity Gradient in the lateral
direction for different  immersion oils , trapping medium is air.}
\end{figure}

\begin{table}[ht!]
\begin{center}
\caption{The optimal trapping depth ($d_{opt}$), equivalent probe
depth ($d$; distance traveled by the objective) and the focus
shift ($\Delta f$) for trapping inside water using different
immersion oils.}
\begin{tabular}{c c c c c c c c c c c}
$n_{im}$                            & 1.518 & 1.53  & 1.54   & 1.55    & 1.56    & 1.57   & 1.58    & 1.59 &1.60 \\
\hline
$d_{opt}(\mu m)$                    & 0     & 3.91  & 7.31   & 9.9    & 12.85   & 15.37   & 17.77   & 19.83 & 26.49 \\
\hline
$d (\mu m)$                         & 0     & 4.5   & 8.5    & 11.5     & 15.0    & 18.0  & 21.0    & 23.5 & 32.5 \\
\hline
$\Delta f(\mu m)$                   & 0     & 0.59  & 1.19   & 1.60    & 2.15    & 2.63    & 3.23    & 3.67 & 6.01\\
\hline $ d_{opt},lateral\quad direction$   & 0     & 4.28  & 7.67
& 10.62    & 13.50    & 16.44    & 19.19 &24.88 &27.21\\
\hline $d(\mu m),lateral\quad direction$   & 0     & 5.0  & 9.0 &
12.50 & 16.0   & 19.5    & 23.0 &30.5 &33.5
\end{tabular}
\end{center}
\end{table}

It can be deduced from table 1 that $0.01$ increment in the
refractive index of the  immersion medium results in $3-3.5 \mu m$
shift for the optimal trapping depth which is in very good
agreement with the previously reported experimental
results\cite{Reihani-oil}. To estimate of the optimal conditions
for the lateral trap, same calculations were repeated for the
lateral intensity distributions. Figure 2, as an example, shows
the lateral intensity distributions at different depths for
$n_{imm}=1.518$ and $n_{imm}=1.58$.

\subsection{Trapping in Air}
Optical tweezers have been used for trapping micro-objects in the
air\cite{Mcgloin}. Same  calculation can be repeated using $m=4$
and $n_{sample}=1$ to find the optimal conditions for the aerosol
trapping. It is worth mentioning that in this case, the total
internal reflection may occur at the coverglass-air interface for
some incident angles (marginal rays). Therefore, there would be an
upper limit for the numerical aperture. Figure 3 shows typical
intensity distributions in the axial directions using 3 different
immersion oils.

\begin{table}[ht!]
\begin{center}
\caption{The optimal trapping depth ($z_{opt}$), equivalent probe
depth ($d$; distance traveled by the objective) and the focus
shift ($\Delta f$) for trapping inside Air using different
immersion oils.}
\begin{tabular}{c c c c c c c c c c c}
$n_{im}$                            & 1.518 & 1.53  & 1.54   & 1.55    & 1.56    & 1.57   &  1.58  &  1.59  &  1.60  &  1.70      \\
\hline
$d_{opt}(\mu m)$                    &  0     & 0.31  & 0.55   & 0.87    & 1.13   & 1.31   & 1.61 & 1.86 & 2.02 & 4.07  \\
\hline
$d (\mu m)$                         & 0         & 0.6   & 1.0    & 1.6     & 2.0    & 2.2 & 2.8 & 3.2 & 3.4 & 6.4  \\
\hline
$\Delta f(\mu m)$                   & 0     & 0.29  & 0.45   & 0.73    & 0.87    & 0.89    & 1.19 & 1.34 & 1.38 & 2.33 \\
\hline $ d_{opt},lateral\quad direction$   & 0     & 0.4  & 0.64 &
0.95    & 1.22    & 1.57    & 1.78 & 2.03 & 2.37 & 4.9
\\ \hline $d(\mu m),lateral\quad direction$   & 0     & 0.8  &
1.2     & 1.8     & 2.2   & 2.8    & 3.2 & 3.6 & 4.2 & 8.4
\end{tabular}
\end{center}
\end{table}

Table 2 summarizes the calculation results for trapping in the
air. It can be deduced  that $0.01$ increment in the refractive
index of the immersion medium shifts the optimal trapping depth by
$0.3\mu m$ which is very small compared to the previous case.
Therefore, to get a reasonable shift in the optimal trapping depth
higher refractive index liquid is required .

\end{document}